\RequirePackage{fix-cm}
\documentclass[smallextended]{svjour3}       
\smartqed  
\usepackage{graphicx}
\usepackage{amsmath}
\usepackage{amssymb}
\usepackage{cite}
\usepackage{textcomp}
\usepackage{graphicx}
\usepackage{physics}
\usepackage{float}
\usepackage{upgreek}
\usepackage{stackengine}
\newcommand\xrowht[2][0]{\addstackgap[.5\dimexpr#2\relax]{\vphantom{#1}}}
\usepackage[dvipsnames]{xcolor}
\usepackage{soul}
\usepackage{ulem} 		
\begin{document}

\title{Vantablack Shielding of Superconducting Qubit Systems}
\titlerunning{Vantablack Shielding of Superconducting Qubit Systems} 

\author{J.M.~Kitzman \and J.R.~Lane \and T.~Stefanski \and N.R.~Beysengulov \and D.~Tan \and K.W.~Murch \and J.~Pollanen}

\institute{J.M.~Kitzman, J.R.~Lane, T. ~Stefanski, N.R.~Beysengulov, J.~Pollanen \at
              Department of Physics and Astronomy, Michigan State University, East Lansing, Michigan, USA \\
              \email{kitzmanj@msu.edu}  
           \and
           D.~Tan, K.W.~Murch \at
              Department of Physics, Washington University, St. Louis, Missouri, USA \\
              Center for Quantum Sensors, Washington University, St. Louis, Missouri, USA
}

\date{}
\maketitle

\begin{abstract}
Circuit quantum electrodynamics (cQED) experiments on superconducting qubit systems typically employ radiation shields coated in photon absorbing materials to achieve high qubit coherence and low microwave resonator losses. In this work, we present preliminary results on the performance of Vantablack as a novel infrared (IR) shielding material for cQED systems. We compare the coherence properties and residual excited state population (or effective qubit temperature) of a single-junction transmon qubit housed in a shield coated with a standard epoxy-based IR absorbing material, i.e. Berkeley Black, to the coherence and effective temperature of the same qubit in a shield coated in Vantablack. Based on a statistical analysis of multiple qubit coherence measurements we find that the performance of the radiation shield coated with Vantablack is comparable in performance to the standard coating. However, we find that in the Vantablack coated shield the qubit has a higher effective temperature. These results indicate that improvements are likely required to optimize the performance of Vantablack as an IR shielding material for superconducting qubit experiments and we discuss possible routes for such improvements. Finally we describe possible future experiments to more precisely quantify the performance of Vantablack to improve the coherences of more complex cQED systems.
\end{abstract}

\section{Introduction}
Superconducting qubits are among the state of the art technologies being developed in the pursuit of a functional quantum computer~\cite{Arute2019,Kandala2019,Kjaergaard2020}, with current depolarization and dephasing coherence times for transmon-based processors exceeding $100~\textrm{$\upmu $s}$~\cite{PRXQuantum.2.020306}, and recent experiments employing heavy fluxonium~\cite{Zhang2021} demonstrating coherence times up to $1~\textrm{ms}$~\cite{somoroff2021millisecond}. One of the primary sources of decoherence in modern superconducting processors is the presence of non-equilibrium quasiparticles (broken Cooper-pairs) in the in the superconducting electrodes that form the circuit~\cite{PhysRevB.74.064515,PhysRevLett.103.097002,PhysRevLett.114.240501}. It is well known that stray black-body infrared (IR) radiation, which has energy greater than the superconducting gap, can break Cooper pairs in the superconductor, increasing the density of non-equilibrium quasiparticles, and poisoning coherence~\cite{doi:10.1063/1.3638063,doi:10.1063/1.3658630,PhysRevApplied.12.014052}. Additionally, it has been found that effectively shielding the system from these IR photons improves in the internal quality factor of the superconducting resonators employed for qubit control and readout~\cite{Kreikebaum_2016}. In order to further increase qubit coherence times it is important to continue to investigate new methods and materials for shielding superconducting qubits and resonators from unwanted sources of decoherence. In this work, we investigate the performance of a new IR shielding material: \textit{Vantablack}. Our preliminary results indicate that Vantablack has potential to yield performance beyond standard coatings when properly applied.

\section{Experimental Details}
 Vantablack is a coating composed of vertically aligned nanotube arrays, grown via a modified chemical vapor deposition process, and exclusively developed by {S}urrey {N}anoSystems {L}td~\cite{SurreyNano}. It is one of the darkest substances known, reflecting less than $0.2\%$ of light in the visible spectrum. These extraordinary absorbing characteristics extend into the infrared spectrum, where Vantablack reflects less than $0.5\%$ of IR photons~\cite{vantablack}. Vantablack is found in many light absorbing applications, including commercial thermal imaging systems\cite{vb} as well as beam dumps for high power optical experiments\cite{10.1117/1.OE.59.5.056108}. 
 
 To investigate the performance of this shielding material in the context of superconducting qubit systems we measure the coherence properties of a single-junction transmon qubit housed in a three-dimensional (3D) microwave cavity~\cite{PhysRevA.76.042319}. In cQED experiments, such as the ones we perform, it is common to enclose the experimental setup in a shielding material that protects both the control/readout resonator and qubit from stray black-body radiation. In our experiments, this is achieved by housing the qubit and resonator in a copper cylinder coated internally with the shielding material, and then thermally anchoring the cylinder to the mixing chamber of a dilution refrigeratorhaving a base temperature of $\simeq 10$~mK, which is much less than the effective temperature associated with the qubit transition frequency (approximately 250~mK). To compare the performance of Vantablack to a standard epoxy-based coating, which serves as a control experiment, we cover the inside of two identical copper cylinders in either a standard IR coating or Vantablack as shown in Fig.~\ref{fig1}a. The standard infrared absorbing coating we use for our control experiment is composed of a mixture of (by mass) $68\%$ Stycast 2850FT epoxy, $5\%$ Catalyst 24LV, $7\%$ carbon lamp black, and $20\%$ 175~$\upmu$m diameter glass beads. This particular type of epoxy-based coating is often referred to as \textit{Berkeley Black}~\cite{Persky1999} and has been used in previous microwave circuit experiments to produce systematic improvements in the quality factor of superconducting resonators~\cite{Kreikebaum_2016}. Then, on two-separate cool-downs of the dilution refrigerator, we place the same microwave cavity and qubit into one of these two copper cylinder and measure the coherence properties of the qubit.

 Both the control and readout of the qubit, which has a frequency $\omega_q/2 \pi=5.165~\textrm{GHz}$, are mediated via the electric field of the TE101 mode of the 3D electromagnetic resonator with frequency $6.936$~GHz (see Fig \ref{fig1}b.)~\cite{PhysRevLett.107.240501}. For a given measurement, microwave pulses of appropriate length and amplitude are used to manipulate the state of the qubit. After a specified amount of free-evolution, the state of the qubit is then inferred via measurement of the transmission through the microwave cavity, which is dispersively shifted by the presence of the qubit~\cite{PhysRevA.69.062320}. In particular, we employ a high-fidelity readout protocol based on the non-linearity of this interaction~\cite{PhysRevLett.105.100505,PhysRevLett.105.100504,PhysRevLett.105.173601}. 
 A more detailed description of the relevant microwave circuit and qubit measurement protocols can be found in Ref.~\cite{PhysRevA.101.012336}.
\begin{figure}[h]
\centering
\includegraphics[width=0.8\textwidth]{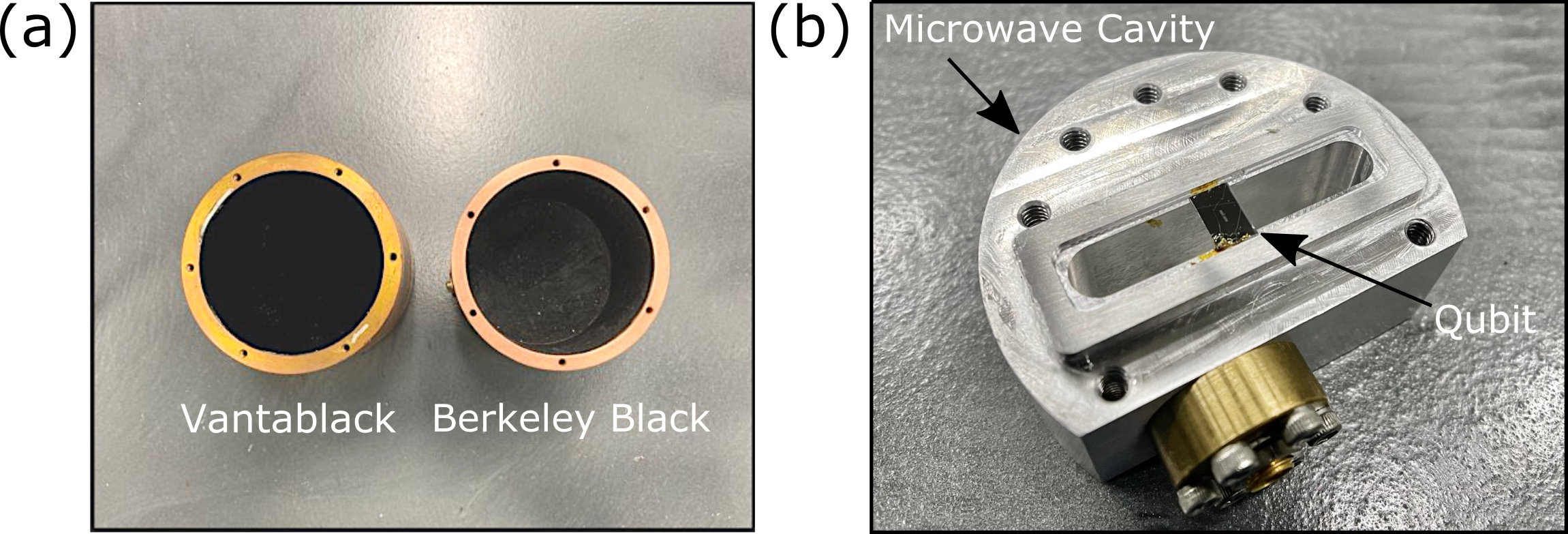}
\caption{\label{fig1}(a) Cylindrical copper housings containing the 3D microwave cavity and superconducting qubit. The interior of these cylinders is coated with either a Vantablack (left) or a Berkeley Black (right) coating. (b) Bottom half of the 3D microwave control/readout cavity and transmon qubit, which was fabricated on high resistivity silicon.}
\label{fig1}
\end{figure}
 

\section{Results and Discussion}
\begin{figure}[H]
\centering
\includegraphics[width=0.8\textwidth]{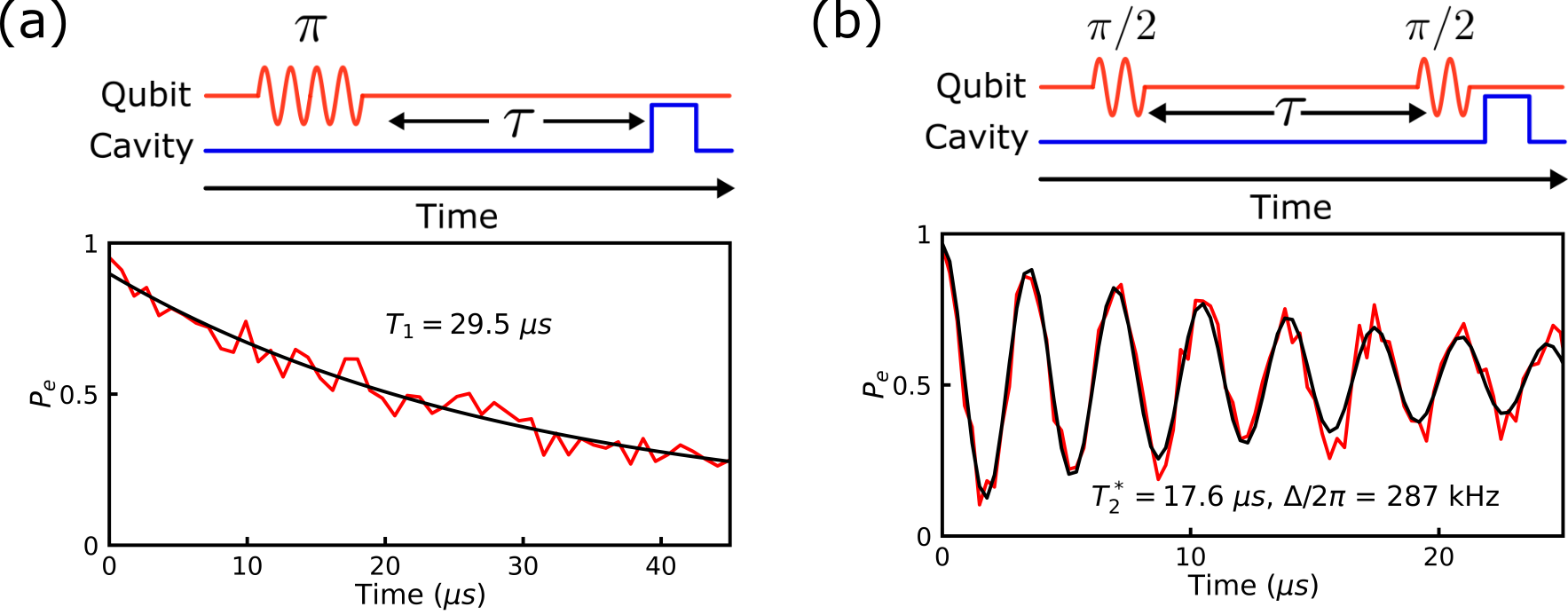}
\caption{\label{fig2} (a) Microwave pulse sequence to measure $T_1$ (top) along with a representative measurement (bottom). (b) Microwave pulse sequence to measure $T_{2}^*$ (top) and a representative measurement (bottom). As described in the text, a detuning $\Delta/2\pi=287$~kHz separated the drive frequency and $\omega_{q}$ in these measurements.}
\end{figure}

To characterize the effectiveness of the coatings at mitigating loss, we measure the depolarization time $T_1$ and the dephasing (Ramsey) time $T_{2}^*$ of the qubit in the copper housings covered in Berkeley Black and Vantablack. In Fig.~\ref{fig2} we show representative measurements of both $T_1$ and $T_{2}^*$ for the qubit housed in the Vantablack coated shield as well as the accompanying microwave pulse sequences applied to perform these measurements. 
Because the measurement of the Ramsey decay time is a phase sensitive measurement, a detuning $\Delta$ between the drive frequency and $\omega_{q}$ will lead to a decaying sinusoid rather than a decaying exponential function. This allows us to extract the magnitude of the detuning between the drive frequency and the qubit frequency in addition to the dephasing time of the qubit.

\begin{figure}[h]
\centering
\includegraphics[width=0.8\textwidth]{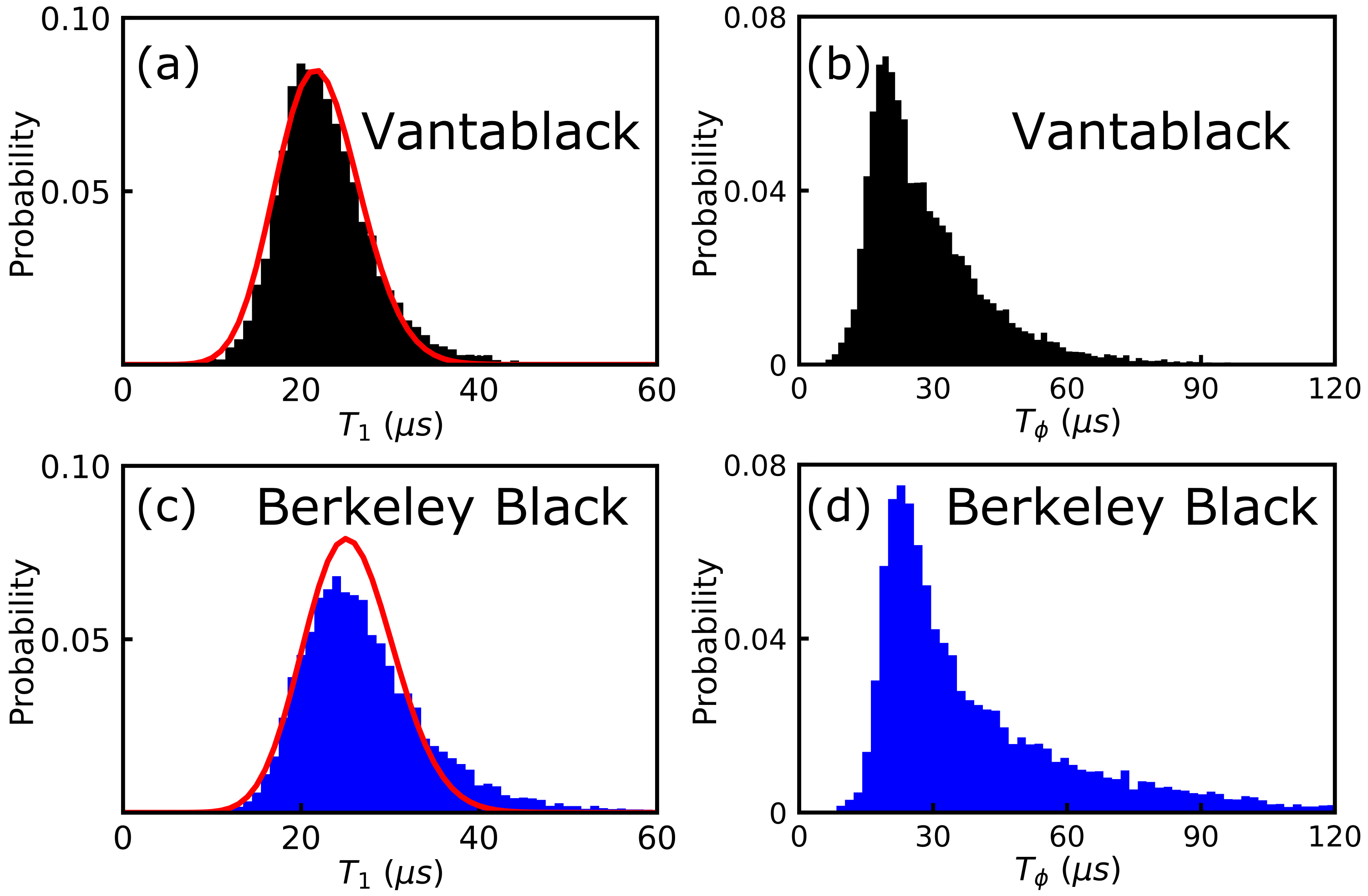}
\caption{\label{fig3} Probability distributions of both $T_1$, along with fits to Poisson distributions (red curves), and $T_{\phi}$ in either the Vantablack coating (a,b) or the Berkeley Black coating (c,d).}
\end{figure}

It is well known that qubit coherences can fluctuate over long timescales due to interactions with quantum two level systems~\cite{PhysRevLett.121.090502,Burnett2019,PhysRevLett.123.190502,PhysRevB.92.035442,PhysRevLett.94.127002} as well as stray radiation from sources such as cosmic rays~\cite{Wilen2021,IBM2020}. Therefore in order to quantify qubit decoherence, we perform repeated measurements of both $T_1$ and $T_2^*$ to obtain sufficient statistics to understand the performance of the two coatings relative to one another. Specifically, we interleave $T_1$ and $T_2^*$ measurements over a 16 hour period and each measurement contains $252$ points in time with $100$ averages at a repetition rate of $9$ kHz. This allows us to extract $T_1$ and $T_{2}^*$ at a rate of approximately $(5~\textrm{s})^{-1}$. From these measurements we extract the pure dephasing time of the qubit,
\begin{equation}
    T_{\phi} = \left(\frac{1}{T_{2}^*} - \frac{1}{2T_1}\right)^{-1}
\end{equation}
\noindent 
and compare the resulting distributions between the two experiments with different shielding materials. The measured probability distributions for both $T_1$ and $T_{\phi}$ for the qubit system shielded by Vantablack and Berkeley Black coatings are plotted in Fig.~\ref{fig3}. In order to compare probability distributions of these coherence measurements, we divide each histogram bin by the total number of experiments to normalize the histograms to have area equal to one. The average values $\overline{T_1}$ and $\overline{T_{\phi}}$ obtained from each distribution are listed in Table~\ref{table} demonstrating a similar level of coherence in both experiments and indicate that Vantablack is a compatible coating with cQED experiments.

\begin{table}[h]
\begin{center}
 \begin{tabular}{||c c c c||} 
 
 \hline\xrowht{8pt}
 Coating & $\overline{T_1}$ ($\mu$s) & $\overline{T_{\phi}}$ ($\mu$s)  & $P_e(\%)$\\
 \hline\hline
 Vantablack & $22.9 \pm 5.6$ & $29.7 \pm 14.9$ & $3.2 \pm 0.59$ \\
 \hline
 Berkeley Black & $27.5 \pm 8.3$ & $45.9 \pm 41.2$ & $0.5 \pm 0.17$\\
 \hline
\end{tabular}
\caption{Averaged results from the qubit coherence distributions shown in Fig.~\ref{fig3} and their standard deviations. $P_{e}$ represents the thermal population of the excited state of the qubit as described in the main text.}
\label{table}
\end{center}
\end{table}

While there does not seem to be a large discrepancy between $\overline{T_1}$ in either shielding environment, it is noteworthy that the average depolarization time in the Berkeley Black environment is slightly longer. Uncorrelated losses of qubit excitations will manifest as a Poissonian distribution in $T_1$~\cite{PhysRevLett.113.247001}, which we find in both sets of measurements regardless of coating. Because the distribution of $T_1$ measurements is governed by Poisson statistics, a larger mean will also lead to a skewed distribution towards higher values of $T_1$, which is consistent with the slightly higher value observed in experiments using Berkeley Black as a coating.

In order to further investigate the differences in performance between these two coatings we measure the residual excited state population of the qubit using a method described in Refs.~\cite{PhysRevLett.110.120501,PhysRevLett.114.240501}, which can be interpreted as an effective qubit temperature. As shown in in final column of Table~\ref{table} we find that the qubit system housed in the Vantablack coating has a residual excited state population of $3.2\%$ as compared to $0.5\%$ in the Berkeley Black coating. This maps onto an effective qubit temperature of $72.6~$mK (Vantablack) and $46.6~$mK (Berkeley Black). Using the principle of detailed balance, we are able to calculated the rate at which spurious excitations in the system drive the qubit from its ground to excited state,
\begin{equation}
    P_g\Gamma_{\uparrow} = P_e\Gamma_{\downarrow},
\end{equation}
where $P_g$ and $P_e$ are the residual population of the qubit ground and excited states, respectively. $\Gamma_{\downarrow} = 1/\overline{T_1}$, and $\Gamma_{\uparrow}$ is the rate at which the qubit is driven from the ground to excited state. We find that in Vantablack, $\Gamma_{\uparrow} \simeq 1.4 \times 10^3 /\text{s}$, and in the Berkeley Black coating, we find that $\Gamma_{\uparrow} \simeq 0.19 \times 10^3 /\text{s}$. Two primary mechanisms may be responsible for the difference in spurious excitation rate between the two coatings. First, the qubit temperature between the two experiments could differ slightly, perhaps due to a relatively low thermal conductivity for Vantablack versus Berkeley Black at mK temperatures. In order to test such a hypothesis, more extensive studies of the thermal properties of Vantablack are necessary. Alternatively, and more interestingly, the density of stray photons present in the Vantablack shielded environment could be higher than in the housing coated in Berkeley Black, indicating that the performance of Vantablack is inferior to that of Berkeley Black. These preliminary measurements do not allow us to unambiguously disentangle these two possibilities and future experiments optimizing the Vantablack coating are needed to identify which may be causing the increased rate of qubit excitation.

In particular, although Vantablack has an extremely low reflectivity at wavelengths in the IR range, it is known that increasing the thickness and roughness of light-shielding coatings can further decrease the reflectivity of a coating. In particular, Ref.~\cite{Persky1999} reports a $\sim 30\%$ reduction in IR reflectance upon increasing the thickness of commercially available coating (Chemglaze Z306) from $25~\upmu$m to $100~\upmu$m. Vantablack is created as a relatively thin coating, with thickness ranging from 20~$\upmu$m to 50~$\upmu$m~\cite{vantablack}. Assuming a similar scaling with thickness would reduce the IR reflectance of Vantablack from 0.5\% to 0.35\%. For comparison Berkeley Black has a typical application thickness near 100~$\upmu$m, or more, with the addition of glass beads which provide extra scattering sites for stray photons. This leads us to believe that the addition of extra materials into Vantablack which increase the scattering of stray photons may improve qubit performance in future experiments using Vantablack as a IR shielding material.

\section{Conclusion}
In conclusion, we have investigated of the utility of Vantablack as a novel shielding material on the coherence properties of a superconducting qubit system. We find that Vantablack does not significantly negatively impact the measured coherence properties of the qubit. It could, however, lead to a higher effective qubit temperature and therefore higher rate of spurious qubit excitation. However, future experiments are needed to more completely understand the potential of this material in improving the state-of-the-art. In particular, future experiments in which high-frequency microwave and IR radiation can be controllably injected are needed to systematically study qubit coherence as a function incident power~\cite{doi:10.1063/1.3638063}. Similarly, planar superconducting qubit geometries have a much smaller mode volume, and are much more sensitive to surface losses~\cite{PhysRevLett.107.240501}, and may respond more dramatically to changes in the IR shielding environment. Building upon the initial experiments reported here, these future experiments can advance the understand of Vantablack as an IR shielding material for cQED systems based in superconducting circuits.

\begin{acknowledgements}
We thank M.I.~Dykman, N.O.~Birge, E.M.~Levenson-Falk, H.~Byeon, L.~Zhang, C.A.~Mikolas, and B.~Arnold for fruitful and helpful discussions. We also thank R.~Loloee and B.~Bi for technical support and use of the W.M. Keck Microfabrication Facility at MSU. N.R.B. acknowledges support from a sponsored research grant from EeroQ Corp. J. Pollanen, J.R. Lane and J.M. Kitzman acknowledge support from the National Science Foundation via grant number DMR-2003815 as well as the valuable support of the Cowen Family Endowment at MSU. K.W.~Murch acknowledges support from the National Science Foundation via grant number PHY-1752844 (CAREER). The data that support the findings of this study are available from the corresponding authors upon reasonable request.

\end{acknowledgements}

\bibliographystyle{spphys}       

%
%

\end{document}